\documentclass[twocolumn,aps,superscriptaddress,prl,10pt]{revtex4-2}
\usepackage{multirow}
\usepackage{graphicx}
\usepackage{dcolumn}
\usepackage{bm}
\usepackage{bbold}
\usepackage{braket}
\usepackage{amssymb,amsmath}
\usepackage[section]{placeins}

\usepackage{color}
\usepackage{amsfonts}
\usepackage{subfigure}
\usepackage{array}
\usepackage{enumerate}
\usepackage{cancel,soul,ulem}
\renewcommand{\sout}{\bgroup \color{red} \ULdepth=-.5ex \ULset}
\usepackage{xspace}
\usepackage{siunitx}
\usepackage{xfrac}
\usepackage{hyperref}
\usepackage[nameinlink]{cleveref}
\usepackage{appendix}
\usepackage{subfigure}
\usepackage{xifthen}
\usepackage{xcolor}
\hypersetup{
	colorlinks,
	linkcolor={red!75!black},
	citecolor={blue!75!black},
	urlcolor={blue!75!black}
}
\graphicspath{{./figures/}{./}}
\begin{document}
\title{Primordial Quark Stars Made of Long-lived False Vacuum}
\author{Jingdong Shao}
\email[]{shaojingdong19@mails.ucas.ac.cn} 
\affiliation{School of Physical sciences, University of Chinese Academy of Sciences, Beijing 100049, China} 
\author{Mei Huang}
\email[]{huangmei@ucas.ac.cn~(corresponding author)}
\affiliation{School of Nuclear Science and Technology, University of Chinese Academy of Sciences, Beijing 100049, China}	
\begin{abstract}
A false vacuum could be a profound ingredient of fundamental physics, yet its direct detection in laboratories is hindered when the lifetime is exponentially long. Conventional static phase diagrams often discard metastable false vacuum, we show that, however, in a dynamical treatment of a first-order QCD phase transition at large quark chemical potential that strongly suppresses tunneling, cosmologically long-lived, and thus indispensable false vacuum naturally arises, and makes nontrivial contribution to constituting primordial quark objects. We identify two distinct branches of such primordial quark objects: quark star crusted with false vacuum, and nugget specifically referring to quark star in global false vacuum, which may account for a population of small compact stars. Some long-lived primordial objects may have survived until the late Universe, and hopefully sow the seeds of high red-shift galaxies. Its false vacuum decay can power ultra-energetic long $\gamma$-ray bursts and kHz gravitational waves within multi-messenger facilities, rendering itself an astrophysical and thus testable phenomenon.
\end{abstract}

\maketitle
The existence of a metastable, or ``false'', vacuum is a fundamental concept in quantum field theory and in many-body physics \cite{Turner:1982iov,Langer:1969bc}. A false vacuum is a local minimum of the effective potential separated from the true ground state by an energy barrier; its decay requires either quantum tunneling or thermal nucleation \cite{RN8,RN28} and is therefore governed by intrinsically nonperturbative dynamics. This mechanism underlies paradigmatic phenomena across physics, from long-lived metastable phases in many-body systems \cite{PMID:38844460,Zenesini:2023afv,Darbha:2024srr,Braden:2017add} to vacuum decay in particle physics and cosmology \cite{Markkanen:2018pdo,Vilenkin:1983boi}. Conventional, equilibrium phase diagrams constructed from static energy minima often discard metastable false vacuum. In contrast, a cosmic first-order transition is a genuinely dynamical process with supercooling and nucleation. When nucleation is strongly suppressed, a region that falls out of equilibrium can become trapped in a metastable state for cosmological timescale. Such long-lived false vacuum would behave as genuine forms of matter and therefore should not be discarded. In this work we show exemplarily that macroscopic volumes of quark matter are trapped as false vacuum for cosmological timescale in a dynamical first-order quantum chromodynamics (QCD) phase transition, and thus can make nontrivial contribution to constituting primordial quark objects. 

Direct detection of a false vacuum is notoriously difficult when spontaneous decay requires exponentially long waiting times that place decay far beyond laboratory timescales. To make vacuum decay experimentally accessible, several groups have proposed engineered probes such as quantum simulation platforms \cite{Vodeb:2024tvo,Lagnese:2023xjg,Zhu:2024dvz,Ng:2020pxk,Abel:2020qzm} and cold atom simulators \cite{Braden:2017add,Jenkins:2023npg,Jenkins:2023eez,Jenkins:2025szl} that emulate aspects of nucleation and tunneling in controllable environments. An alternative, complementary possibility is that the early Universe itself produced macroscopic regions of metastable matter whose eventual decay would release large, localized amounts of energy and thereby convert an otherwise inaccessible quantum process into an astrophysical observable.

We show that such primordial ``decay seeds'' naturally arise from primordial quark objects in the first-order QCD phase transition at high quark chemical potential $\mu$. 
In a locally high $\mu$ region generated by inhomogeneous baryogenesis mechanisms \cite{Affleck:1984fy,Dine:2003ax,Dolgov:2008wu,Linde:1985gh,Yokoyama:1987he,Kartavtsev:2008fp} or inhomogeneity formation mechanisms \cite{Byrnes:2018txb,Mishra:2019pzq,Ballesteros:2021fsp,Lin:2020goi,Fu:2022ypp,Qiu:2022klm,liu2022primordial,Zhang:2025kbu,Jedamzik:1996mr,Kanemura:2024pae}, QCD phase transitions are of first-order \cite{PhysRevD.73.014019,Bao:2024glw,Wang:2023pmn,Wang:2023omt}. Crucially, at sufficiently large $\mu$ the effective potential is modified so that nucleation is further inhibited, and a critical nucleation point \cite{Shao2024short,Shaojd2024long}, at which the nucleation can barely start, is found. Quark matter is therefore trapped as false vacuum and meanwhile forms celestial objects in these locally dense region. Consequently, the false vacuum hopefully decays during the long time since the early Universe as a natural source. 

High red-shift galaxy candidates observed by the James Webb Space Telescope (JWST) strongly suggests the existence of locally high baryon density in the early universe other than uniform low density. Abundant early galaxies need a rapid galaxy formation and evolution mechanism \cite{Tacchella_2018} different from the standard $\Lambda$CDM cosmology \cite{Boylan-Kolchin:2022kae}. The primordial quark objects in the locally dense region may have survived until the late Universe due to extremely slow false vacuum decay rates and may sow the seeds of early galaxies, then subsequent accretion and release of immense vacuum energy can further facilitate star formations.

Primordial quark object itself also provides direct and testable implications. First, they constitute a class of compact celestial bodies analogous to strange stars \cite{Alcock1986,Haensel1986,Dey1998} and possess unique false vacuum constituents based on quark star models \cite{Yuan:2022dxb,Yuan:2025yjy}. They may account for a population of compact stars with radii and masses smaller than typical hadronic neutron stars. Typical compact objects including HESS J1731$-$347 \cite{Victor:2022asl,Horvath:2023uwl,Sagun:2023rzp} and others \cite{Dey1998,Li2015_4U1746,Li1999_SAXJ1808,Xu2005_1E1207,Gangopadhyay2013,Bagchi:2008ki} have been discussed as potential quark-star candidates. Observational efforts have revealed more small compact stars as potential candidates, including low-mass X-ray binaries (LMXBs) 4U~1820$-$30 \cite{stella1987discovery,strohmayer2002remarkable,Guver2010_4U1820}, 4U~1608$-$52 \cite{belian1976discovery,Guver2008_4U1608}, 4U~1746$-$37 \cite{1977IAUC.3095,Li2015_4U1746}, SAX~J1750.8$-$2900 \cite{natalucci1999new,Lowell2012_SAXJ1750}, EXO~1745$-$248 \cite{makishima1981discovery,Ozel2009_EXO1745} and SAX~J1748.9$-$2021 \cite{intZand:1999blh,Guver2013_SAXJ1748} from the Rossi X-ray Timing Explorer (RXTE) \cite{galloway2008thermonuclear}, quiescent LMXBs (qLMXBs) U24 in NGC~6397 and X5 in 47~Tuc \cite{grindlay2001chandra,Guillot2011_U24,heinke2003analysis,Bogdanov2016_47Tuc}, the rapid burster MXB~1730$-$335 \cite{lewin1976discovery,Sala2012_RapidBurster}, recent J0437$-$4715 \cite{johnston1993discovery,Choudhury:2024xbk} and J0740$+$6620 \cite{riley2021nicer,Salmi:2024aum} from the Neutron Star Interior Composition Explorer (NICER), examples \cite{Reardon:2015kba} from the Parkes Pulsar Timing Array (PPTA), binaries associated with gravitational wave (GW) events such as GW170817 \cite{LIGOScientific:2018cki,LIGOScientific:2017vwq}.

Second, when false vacuum in such primordial objects eventually decays to the true vacuum, it releases vacuum energy that would be efficient to produce surface eruptions and associated electromagnetic (EM) signals \cite{Drago2015_TwoFamilies,Drago:2015dea}. A number of EM transients, from extremely energetic long $\gamma$–ray burst (GRB) with isotropic energy $E_{iso}\sim10^{55}$ erg \cite{Frederiks:2023bxg,tajima2009fermi} to the giant soft GRB repeater flare of SGR~1806$-$20 \cite{Palmer2005_SGR1806,Gaensler2005_radio} with $E_{iso}\sim10^{46}$ erg, have demonstrated that stellar surfaces can undergo catastrophic and short energy-release events. The immense isotropic energy in long GRB events, which is beyond the upper limit of magnetar engine \cite{Yu:2017fsg}, requires new powerful engine. Independently, a first–order phase transition \cite{Herzog:2011sn}, or even violent quark nova \cite{Ouyed2001_QuarkNova,Ouyed2022_Macro}, can drive the combustion inside a compact star and the release of $10^{53}$ erg of vacuum energy makes quark stars attractive candidate sources of ultra-relativistic outflows and extremely energetic long GRBs \cite{Bombaci:2000cv,Song:2025zdn,Paczynski:2005nt}.

Third, bubble dynamics after false vacuum decay generates GW signals. In compact stars phase transitions can influence the spectra of GW from mergers \cite{Bauswein2019_PRL,Grippa:2024ppo,Mannarelli:2018pjb,bauswein2010discriminating} and drive GWs with high frequencies around $0.1-100$ kHz and the characteristic strain around $h_c\sim 10^{-25}-10^{-19}$ \cite{Lin2006_PhaseTrans,Abdikamalov:2008df,Sigl2006_GWbackground,Mallick:2020bdc,Prasad:2017msy,Prasad:2019kuz}, within the sensitivity band of ground-based GW detectors. Thus primordial quark objects turn the fundamental problem of false vacuum decay into an observational opportunity: long-lived false vacuum objects act as natural decay sources whose signatures can be sought with existing and next-generation multi-messenger facilities.

In this letter, we propose for the first time that, in locally  high chemical potential region, quark matter remains false vacuum and can constitute unique primordial quark objects as detectable false vacuum decay source. We construct the dynamical phase diagram of QCD chiral phase transition with false vacuum window, and consider the lifetimes of different conditions. In the following, we show two different possible destinies of two types of quark objects, quark star with a true vacuum core and a false vacuum crust, and nugget with a global false vacuum. If the phase transition never starts, we argue that both quark objects will be compressed to be quark nuggets with a global true vacuum, similar to primordial quark nuggets (PQNs) proposed by Witten \cite{Witten:1984rs}. If vacuum energy is released, we estimate the potential EM and GW signatures.

\textit{The dynamical phase diagram.}
We employ the two-flavor quark meson (QM) model to describe the chiral phase transition. We denote $\mu_i (i=u,d)$ as chemical potentials of up and down quarks, and $\mu$ as average quark chemical potential. The grand potential \cite{Zhou_2021,Wang:2023omt} is
\begin{equation}
\begin{aligned}
&\Omega=\frac{\lambda}{4}(\sigma^2-v^2)^2-H\sigma-6T\sum_{i=u,d}\int\frac{d^3\vec q }{(2\pi)^3}\left\{ \frac{E_i}{T}+\right.\\
& \left.\left[\mathrm{ln}(1+e^{-(E_i-\mu_i)/T})+\mathrm{ln}(1+e^{-(E_i+\mu_i)/T})\right]\right\},
\end{aligned}
\end{equation}
where $E_i=\sqrt{\vec{q}^2+(g\sigma)^2}$ $(i=u,d)$ is the energy with the coupling constant $g=3.3$ and $\vec q$ is the momentum. We fix $H=f_{\pi}m^2_{\pi}$ by partial conservation of the axial current with $f_\pi=93$ MeV and the pion mass $m_{\pi}=138$ MeV, $\lambda$ by the $\sigma$ mass $m^2_{\sigma}=m^2_{\pi}+2\lambda f^2_{\pi}=(500\mathrm{MeV})^2$, and $V$ by $v^2=f^2_{\pi}-\frac{m^2_{\pi}}{\lambda}$. 

For compact objects, we solve the gap equation $\partial \Omega/\partial \sigma$ with $\beta$-equilibrium and electric charge neutrality. Neutrinos are assumed to escape freely. The electron chemical potential is denoted as $\mu_e$ and the electron mass is negligibly small, thus the pressure of free electron gas is
\begin{equation}
    p_e=2T\int\frac{d^3\vec q}{(2\pi)^3}\left[\mathrm{ln}(1+e^{-\frac{\lvert \vec q \rvert-\mu_e}{T}})+\mathrm{ln}(1+e^{-\frac{\lvert \vec q \rvert+\mu_e}{T}})\right]
\end{equation}
and the total pressure is $p=-\Omega+p_e-p_{vac}$, where $p_{vac}=0.0001064836$ GeV$^4$ is the vacuum pressure in the QM model and must be deducted. The total energy density is 
\begin{equation}
    \rho=-p+T\frac{\partial p}{\partial T}+\mu_un_u+\mu_dn_d+\mu_en_e,
\end{equation}
where $n_i$($i=u,d,e$) denotes the number density of the corresponding particle.

\begin{figure}
    \centering
    \includegraphics[width=0.47\textwidth]{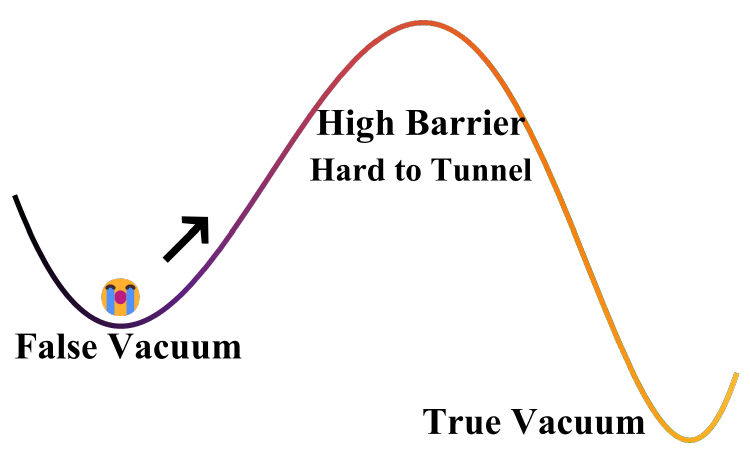}
    \caption{The schematic diagram of the potential with high barrier that suppresses the tunneling process of bubble nucleation in a dynamical first-order phase transition.}
    \label{potential}
\end{figure}

\begin{figure}
    \centering
    \includegraphics[width=0.47\textwidth]{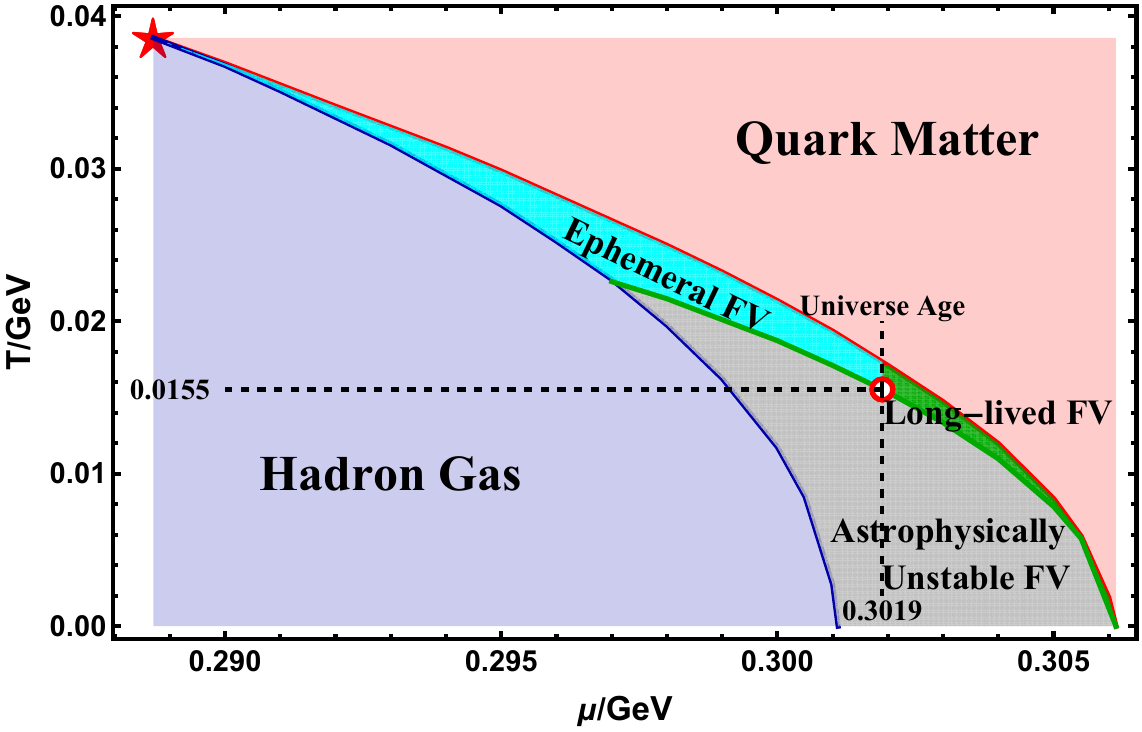}
    \caption{The dynamical phase diagram with the temperature $T$ and the quark chemical potential $\mu$. The red boundary line of the red shaded true vacuum quark matter region is the static first-order phase transition line with the critical end point at $\mu=0.2887$ GeV and $T=0.03859$ GeV indicated by the red star. The blue boundary line of the blue shaded hardon region is the spinodal line, below which no false vacuum quark matter can exist. The green boundary line marks the relation $T(\mu)$ for which the total pressure vanishes. The red ring at $\mu=0.3019$ GeV and $T=0.0155$ GeV indicates the vicinity where the lifetime of $1$ km$^3$ false vacuum is around the age of the Universe. The false vacuum (FV for short in the figure) can be long-lived in the green shaded region but is ephemeral in the cyan region or astrophysically unstable with negative total pressure in the gray region.}
    \label{phase}
\end{figure}
 
We consider dynamically evolving quark matter and take the tunneling process of bubble nucleation into account with gap equations. As we show in Fig. \ref{potential}, the tunneling and bubble nucleation in a dynamical evolution are suppressed if a high barrier is induced. The false vacuum therefore is trapped if the nucleation rate is sufficiently low. The nucleation rate $\Gamma$ per volume and per unit time at finite temperature is
\begin{equation}
    \Gamma=T^4\left(\frac{S_3}{2\pi T}\right)^{\frac{3}{2}} e^{-\frac{S_3}{T}},
\end{equation}
where $S_3$ is the bounce action of the O(3)-symmetric bubble, and we estimate the lifetime of intact false vacuum $\tau$ in a given volume $V$ as the time required for one bubble to nucleate $\tau\sim\frac{1}{\Gamma V}$.

We show the dynamical first-order phase diagram with the temperature $T$ and the quark chemical potential $\mu$ in Fig. \ref{phase}. The red boundary line of the quark matter region represents the static first-order phase transition line with the critical end point (CEP) at $\mu=0.2887$ GeV and $T=0.03859$ GeV indicated by the red star. When the temperature cools down to the critical temperature, the first-order phase transition in the high $\mu$ region does not start immediately, as schematized in Fig. \ref{potential}. The blue boundary line is the spinodal line, at which the potential barrier vanishes and no false vacuum survives.

We further constrain the supercooling by two factors. First, the total surface pressure of a compact star must be positive. The green boundary line marks the relation $T(\mu)$ for which the total pressure vanishes. The region below the green line (shaded gray) therefore has negative total pressure and is astrophysically unstable. Second, the nucleation rate of the false vacuum must be sufficiently low. The red ring at $\mu=0.3019$ GeV and $T=0.0155$ GeV indicates the vicinity where the lifetime of $1$ km$^3$ false vacuum is around the age of the Universe, i.e., around $10^{10}$ years. In addition, false vacuum in the cyan region is ephemeral and will inevitably release the vacuum energy in an extremely short time. By contrast, the long-lived false vacuum in the green region have lifetimes exceeding $10^{10}$ years and therefore tends to persist as temperature cools. 

We expect that, objects, whose boundary values of chemical potential and temperature are around the red ring, have lifetimes close to the age of the Universe and thus are the most likely sources of observations. If we mix quark matter with other positive-pressure matter, we may extend the window of long-lived and stable false vacuum.

\textit{Primordial quark star.} In the following we focus on pure quark objects and consider examples with uniform temperature $T=10$ MeV and $15.5$ MeV. We attain the relation between mass $M$ (in units of the solar mass $M_\odot$) and radius $R$ by solving the Tolman–Oppenheimer–Volkoff equation
\begin{equation}
    \frac{\mathrm{d}p}{\mathrm{d}r}=-\frac{GM}{r^2}(\rho+p)(1+\frac{4\pi r^3 p}{M})(1-\frac{2GM}{r})^{-1},
\end{equation}
where $G$ is the gravitational constant in natural units.

\begin{figure}
    \centering
    \includegraphics[width=0.47\textwidth]{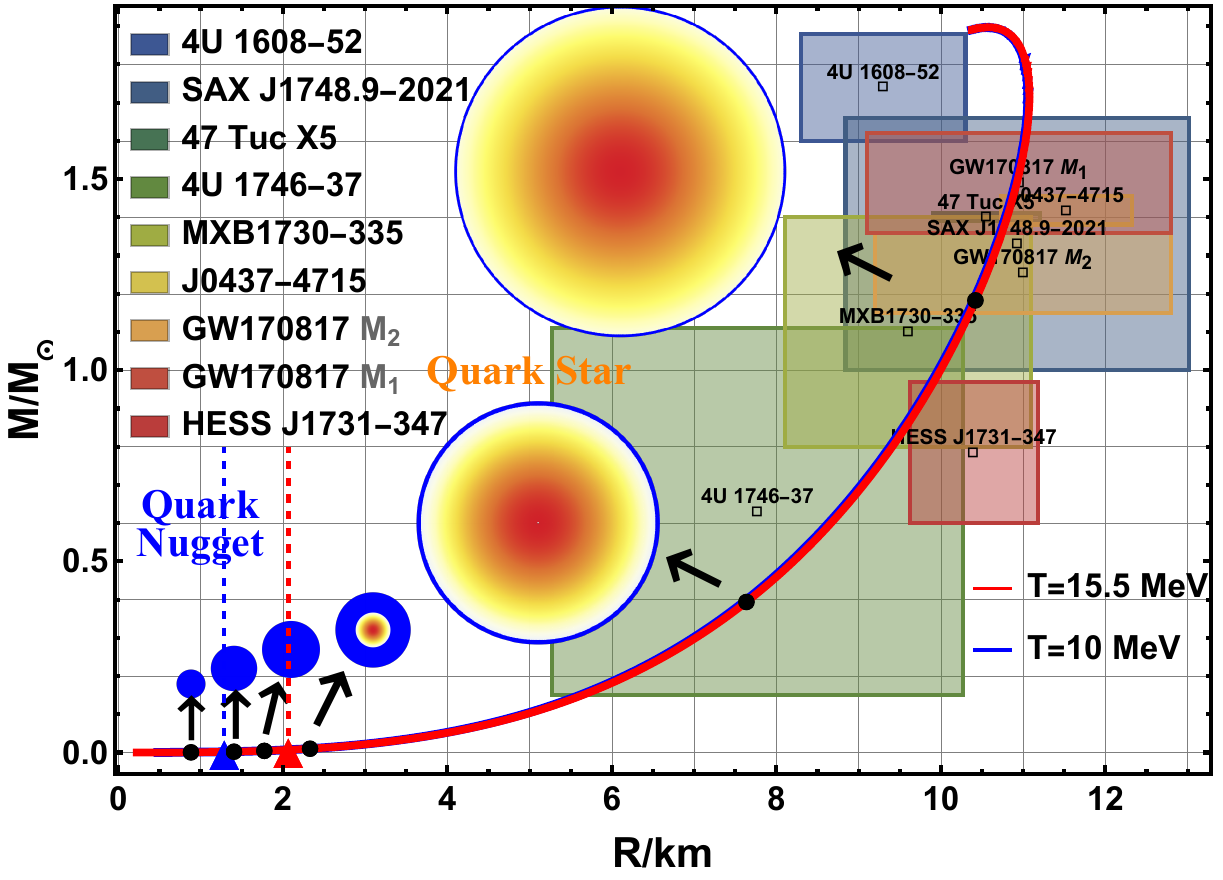}
    \caption{The mass-radius relation of quark objects at $T=15.5$ MeV (red) and $T=10$ MeV (blue). The triangles and dashed lines indicate the boundaries between false-vacuum-crusted stars and nuggets, and six schematic diagrams of representative objects are chosen from the red line (black dots). The shaded regions show the observational constraints from various sources (see text for references).}
    \label{MR}
\end{figure}

We show the mass-radius relation of quark objects at $T=10$ MeV (blue) and $15.5$ MeV (red) together with schematic diagrams of representative objects in Fig. \ref{MR}. The two curves split into two distinct branches by two dashed lines respectively, which correspond to quark stars with true vacuum quark matter core and false vacuum crust (we name them tangyuan stars after this traditional food), and to nuggets specifically referring to quark stars with global false vacuum. The boundary points are at $R=2.069$ km and $M=0.00698M_\odot$ for $T=15.5$ MeV and at $R=1.289$ km and $M=0.00172M_\odot$ for $T=10$ MeV, as indicated by two triangles.

We choose six points from the red line (black dots) and draw exemplary objects in scale. The warm-toned sphere denotes the true vacuum core, with energy density decreasing outward toward the crust. The blue ring denotes the false vacuum layer of a false-vacuum-crusted star, while the blue sphere denotes the global false vacuum of a nugget. With decreasing mass, although sizes of objects monotonically decrease, thickness of false vacuum crust is $\mathcal{O}(10)-\mathcal{O}(100)$ m with $\mathcal{O}(1)M_\odot$, and increases until reach a maximum $\mathcal{O}(1)$ km at the boundary, at which false vacuum occupies whole stars, then false vacuum diminishes with decreasing size. Therefore, objects near the boundary release most false vacuum energy. 

We find that quark stars typically possess $R\sim\mathcal{O}(10)$ km and $M\sim\mathcal{O}(1)M_\odot$ and may explain some of the observational constraints in the diagram, including 4U~1746$-$37 \cite{Li2015_4U1746}, 4U~1608$-$52 \cite{Guver2008_4U1608}, one possible solution for SAX~J1748.9$-$2021 \cite{Guver2013_SAXJ1748}, U24 in NGC~6397 \cite{Guillot2011_U24}, X5 in 47~Tuc \cite{Bogdanov2016_47Tuc}, MXB~1730$-$335 \cite{Sala2012_RapidBurster}, J0437$-$4715 \cite{Choudhury:2024xbk}, HESS J1731$-$347 \cite{Sagun:2023rzp} and the binary in GW170817 \cite{LIGOScientific:2018cki,LIGOScientific:2017vwq}. In the higher temperature case, encrusted stars have radii ranging from $2.069$ km to around $11$ km and masses ranging from $0.00698M_\odot$ to around $1.9M_\odot$, while nuggets occupy the narrow window $M<0.00698M_\odot$ and $R<2.069$ km. In the lower temperature case, although the two curves change very little with temperature, the window for nuggets is much narrower.


We argue that the destiny of false vacuum is set by the competition between the nucleation and compression, quark stars will either be converted into neutron/hybrid stars after nucleation, or remain as stable true vacuum nuggets. First, if the nucleation does not start, as the objects cool, they gradually contract and the surface chemical potential increases to satisfy $p=0$. As temperature approaches $0$, false vacuum in quark stars is compressed into true vacuum with very little change in masses and radii.

Second, we estimate lifetimes of intact false vacuum by nucleation rate on the surface. In the $T=10$ MeV case, the bounce action $S_3/T$ on the surface is around $4403$ and the nucleation rate $\Gamma\sim e^{-4403}$ GeV$^4$ is essentially zero for any size. Therefore, initially cold and isolated quark stars are effectively stable. In the $T=15.5$ MeV case, the bounce action on the surface is around $221.47$, giving the nucleation rate $\Gamma\sim10^{-102}\ \mathrm{GeV}^4=10^{-11}/\ (\mathrm{km}^{3}\times\mathrm{year})$. This rate allows spontaneous phase transition in quark stars whose false vacuum is around $\mathcal{O}(10)\sim\mathcal{O}(100)$ km$^3$, the corresponding lifetimes are $10^{9}\sim10^{10}$ years. Hence initially hot quark stars can spontaneously undergo a first-order phase transition that emits observable signals. Quark stars will become hybrid stars if bubbles are constrained in the crust, or become neutron stars if bubbles spread to the whole object.

If false vacuum undergoes the nucleation, the vacuum energy density is $\epsilon_v=(-\Delta\rho+3\Delta p)/4$, where $\Delta$ means the differences between true and false vacua, and the volume $V$ of false vacuum crust ranges from $0$ to $\mathcal{O}(100)$ km$^3$. The total vacuum energy $E_v=\epsilon_v V$ of the crust can be as large as $10^{52}$ erg at maximum. If the nucleation chain combusts the whole quark star, the immense total vacuum energy can be two orders of magnitude greater than that of the crust. We thus estimate the total vacuum energy with a upper limit of $10^{54}$ erg.

The isotropic energy depends on the efficiency and the geometry
\begin{equation}
    E_{iso}f_b=\eta_j \eta_\gamma E_v,
\end{equation}
where $f_b=1-cos(\theta_j)$ is the beaming factor with $\theta_j$ the jet open angel. $\eta_j$ is the efficiency of converting vacuum energy into jet energy, and $\eta_\gamma$ is the radiative efficiency of converting jet energy into radiation. Efficiencies are estimated to be $10^{-2}\sim10^{-1}$ \cite{Zhang:2006uj,Kumar:2014upa,Panaitescu:2001bx,Narayan:2008xq}, and are corrected by the geometry, i.e., $\theta_j$ that is estimated to be several degrees \cite{Frail:2001qp,Ghirlanda:2004me,Mooley:2018qfh,Troja:2018uns,Fong:2015oha}. For an idealized estimate, we take $E_v\sim10^{54}$ erg and assume that nearly all vacuum energy is converted into the energy of a narrow jet ($\theta_j\sim2^\circ$), the maximum isotropic energy can reach an upper limit of $10^{56}$ erg given $\eta_\gamma\sim0.1$. Thus we predict that this upper limit could in principle power extremely energetic long GRBs ($10^{55}$ erg), and also span the typical GRB energy scale.

Part of the vacuum energy can be emitted as GW bursts, which are dominated by the quadrupole dynamics. A rough estimate gives the frequency $f\sim c_s/R\sim$ kHz and the characteristic strain $h_c\sim \sqrt{G E_{G} /(c^3fd^2)}$ \cite{moore2014gravitational} with $E_{G}$ the total GW energy and $d$ the distance. For example, if GWs carry $10^{51}$ erg of energy like those from phase transitions in neutron stars and the distance from the source beyond the Milky Way is $10$ Mpc, we estimate the characteristic strain as $h_c\sim 10^{-21}$, which is potentially detectable by future ground-based GW detectors.

\textit{Outlook.} We propose that, on the one hand, false vacuum in dynamical cosmic first-order phase transitions is inevitable and should not be discarded, and can make important contribution to constituting unique primordial objects in the Universe. On the other hand, primordial objects are natural laboratories that hold false vacuum since the early Universe and hopefully provide signatures of false vacuum decay. We may apply this scenario in various fields for different unique objects and for detection of different false vacuum, e.g., in nuclear matter with liquid-gas phase transition and condensed matter. One may create long-lived false vacuum on Earth as unique material, and as new reservoirs of energy that can release energy when needed and triggered by proper turbulence. Besides, some models of inflation, electroweak theory, dark sector and so on, involve first-order phase transitions in the early Universe, and one may obtain new interesting results with the dynamical treatment.

Primordial quark objects are natural engines of GRBs and GW bursts without extra mechanisms, especially of extremely energetic long GRBs. Nucleation also modifies signatures in collision and merger events. Such violent events will tear objects apart and heat them to $40$ MeV \cite{PhysRevD.39.1233} and trigger nucleation. The nucleation, either spontaneously or forcibly, starts from the crust and may spread to the whole star. This ``inverse combustion'' process, involves bubble dynamics, curved metric \cite{Isidori:2007vm,Salvio:2016mvj,Garriga:1993fh}, finite volume effects \cite{Fraga:2003mu}, magnetic fields \cite{Kroff:2014qxa}, geometrical shapes, surface effects, rotation, electromagnetic coupling, impurities and so on, worth wide future investigations guided by observations.

We expect that primordial quark objects could serve as seeds of early stars and galaxies. They have already existed since the early Universe and contain vacuum energy with a upper limit $10^{54}$ erg much larger than that in a supernova, the ejecta and shock waves could impact nearby nebulae and trigger early star formation analogous to supernova-induced star birth \cite{herbst1977observational,kumar2010supernovae,herbst1979supernovas}. Remnant objects may also accrete surrounding matter and ignite new stars. Existence of primordial quark objects thus could substantially accelerate the formation of early stars and may explain high red-shift massive galaxies observed by JWST. If primordial quark objects trigger early star formation, we may detect local history of abundance evolution of heavy chemical elements, similar with supernova nucleosynthesis, especially in the first generation of stars (population \uppercase\expandafter{\romannumeral3} stars) \cite{thielemann2003supernova,thielemann2018nucleosynthesis,tominaga2007supernova}. Additionally, primordial quark objects may end collapsing into primordial black holes as seeds of massive black holes. Besides, both outcomes, population \uppercase\expandafter{\romannumeral3} stars \cite{deSouza:2011ea,Campisi:2011qw,toma2011population,Toma:2016wvr} and primordial black holes \cite{genolini2020revisiting,barco2021primordial,Zhilyaev:2007rx}, are possible sources of long GRBs.

We may detect properties of quark matter through thermal history of primordial quark objects. Although nucleation may be more important at high temperatures ($T>20$ MeV) \cite{PhysRevD.39.1233}, the surface evaporation of cold immortal objects may dominate. The evaporation involves the binding energy \cite{Holdom:2017gdc,Yuan:2022dxb,Zhou:2017pha}, the surface tension \cite{PhysRevD.99.014046,Pinto:2012aq} and the initial conditions \cite{Miao:2024qik}, and we may detect theses highly uncertain properties through cooling process. As in the two-flavor case \cite{Miao:2024qik}, we may extend the results considering color superconducting phase \cite{Rajagopal:2000wf,Huang:2004ik,Alford:2007xm,Yuan:2023dxl}, strange quarks or more flavors. Furthermore, we may also extend to other supercooled matter (e.g., self-bound Fermi-balls \cite{Kawana:2022lba}, pions \cite{andersen2009pion,brandt2018new}, nuclear matter) that possibly constitutes celestial bodies, or to primordial black holes made of false vacuum.

Finally, we may constrain the interaction of nuggets or extend them to the dark sector \cite{Bai:2018dxf} so that they can constitute dark matter. ``Dark" nuggets and primordial black holes both as candidates of dark matter are worth astrophysical constraints.

\section{Acknowledgment}
We thank Wenli Yuan, Dianwei Wang and Lei Zhang for helpful discussion. This work is supported in part by the National Natural Science Foundation of China (NSFC) Grant Nos. 12235016, 12221005 and the Fundamental Research Funds for the Central Universities.

\bibliographystyle{unsrt}
\bibliography{main}
\end{document}